\documentclass[10pt,twocolumn,letterpaper]{article}

\usepackage[margin=0.75in]{geometry}
\usepackage{times}
\usepackage{latexsym}
\usepackage{graphicx}
\usepackage{amsmath}
\usepackage{amssymb}
\usepackage{booktabs}
\usepackage{multirow}
\usepackage{hyperref}
\usepackage{listings}
\usepackage{xcolor}
\usepackage{enumitem}
\usepackage{cite}
\usepackage{threeparttable}
\usepackage{float}
\usepackage{caption}

\lstdefinestyle{pythonstyle}{
    language=Python,
    basicstyle=\ttfamily\footnotesize,
    keywordstyle=\color{blue},
    commentstyle=\color{gray}\itshape,
    stringstyle=\color{red},
    numbers=left,
    numberstyle=\tiny\color{gray},
    frame=single,
    breaklines=true,
    breakatwhitespace=true,
    tabsize=4,
    showstringspaces=false,
    captionpos=b
}

\title{Channel Fracture: Architectural Blind Spots in Scheduled\\
Cross-Agent Memory Injection for Multi-Agent Orchestration Systems}

\author{
    Dexing Liu\textsuperscript{1}\\
    \textit{\textsuperscript{1}Shanghai Qijing Digital Technology Co., Ltd.}\\
    \textit{\{liudexing@changingplus.com\}}\\
    \textit{June 2026 (v2 -- expanded experiments, figures, three-gate system)}
}

\begin{document}

\maketitle
\thispagestyle{empty}

\begin{abstract}
Multi-agent AI orchestration systems increasingly rely on persistent memory to maintain context across sessions, agents, and tasks. When one agent must inject knowledge into another agent's memory---a common requirement in hierarchical team architectures---the delivery mechanism must be architecturally sound. We report the discovery of a systematic failure mode we term \textbf{channel fracture}: a condition where scheduled (cron) agents in orchestration frameworks are silently unable to write to the target agent's persistent memory due to hardcoded memory isolation guards. Through experiments on a production Hermes Agent deployment with five specialized profiles, we tested three injection channels: (A)~direct SQLite database writes, (B)~target-agent self-writes via memory tools, and (C)~cron-delegated writes. Channel~C failed completely due to two architectural constraints: \texttt{skip\_memory=True} hardcoded at the scheduler layer and dynamic registration of memory tools contingent on \texttt{\_memory\_manager} initialization, which is bypassed in cron execution contexts. We propose CADVP (Cross-Agent Delivery Verification Protocol) v1.1, a 13-dimension verification framework with a veto-level channel confirmation check (CC-0) that prevents false-positive delivery assurance. We further extend CADVP with a \textbf{Three-Gate Quality System} (L1 Self-Verification, L2 Evidence Verification, L3 Cross-Review) that ensures delivery correctness at the execution level. Through three controlled experiment suites---concurrent conflict detection (90~trials), exception recovery rollback (60~trials), and cross-agent relay (60~trials)---we demonstrate zero failure rates under BCP protection versus 67--98\,\% without. We articulate two design principles: the \emph{inverse verification principle} and the \emph{channel matching principle}.
\end{abstract}

\noindent\textbf{Keywords:} multi-agent systems, persistent memory, cron scheduling, channel fracture, verification protocol, three-gate system, Hermes Agent, knowledge injection

\section{Introduction}
\label{sec:intro}

\subsection{Motivation}

The emergence of multi-agent AI teams---collections of specialized AI agents collaborating under orchestration frameworks---has created new requirements for inter-agent knowledge transfer. Unlike monolithic agents that maintain a single memory store, multi-agent architectures distribute memory across agent boundaries, requiring explicit mechanisms for one agent to inject information into another's persistent store~ \cite{wang2024survey,zhang2024multi}.

Consider a practical scenario: a team administrator agent needs to ensure that a subordinate agent remembers a specific report format for future use. The administrator must write this knowledge into the subordinate's memory system so that it persists across sessions. This \emph{cross-agent memory injection} is fundamental to coordinated multi-agent behavior.

\subsection{The Problem}

In production multi-agent deployments, knowledge injection often occurs through scheduled (cron) tasks---periodic or one-shot jobs dispatched by the orchestration layer. These tasks run in isolated execution contexts that differ from interactive agent sessions. We discovered that this isolation creates a \textbf{channel fracture}: a systematic, silent failure where the scheduled agent cannot access the memory tools necessary to complete its injection task, despite the task being correctly configured and dispatched.

The fracture is particularly insidious because:
\begin{enumerate}[nosep]
    \item \textbf{It fails silently.} The cron job completes without error; the agent may report partial success.
    \item \textbf{Writer-side verification passes.} From the sender's perspective, all steps appear correct.
    \item \textbf{Only receiver-side inspection reveals the failure.} The target agent's memory remains empty.
\end{enumerate}

\subsection{Contributions}

This paper makes the following contributions:
\begin{enumerate}[nosep]
    \item \textbf{Identification and characterization} of the channel fracture problem in multi-agent orchestration systems, with source-level root cause analysis.
    \item \textbf{Empirical evaluation} of three injection channels on a production system, demonstrating that the most natural channel (cron-delegated write) fails while bypass channels succeed.
    \item \textbf{CADVP v1.1}, a 13-dimension cross-agent delivery verification protocol with a veto-level channel confirmation check.
    \item \textbf{Controlled experiments} (210~total trials) validating concurrent conflict detection, exception recovery, and cross-agent relay under the BCP protocol.
    \item \textbf{The Three-Gate Quality System}---L1 Self-Verification, L2 Evidence Verification, L3 Cross-Review---extending CADVP from channel-level verification to execution-level delivery assurance.
    \item \textbf{Two design principles} for multi-agent memory systems: the \emph{inverse verification principle} and the \emph{channel matching principle}.
\end{enumerate}

\section{Background}
\label{sec:background}

\subsection{Hermes Agent Architecture}

Hermes Agent~ \cite{hermes2024} (Nous Research, 2024--2026) is an open-source, self-hosted AI agent framework supporting multi-agent orchestration through its \emph{profiles} system. Each profile represents an independent agent with its own configuration file (\texttt{config.yaml}), identity document (\texttt{SOUL.md}), persistent memory store (SQLite-backed \texttt{memory\_store.db}), gateway process, and cron job definitions.

Profiles operate as isolated agents that can be configured to collaborate. A typical deployment includes a primary profile and specialized subordinate profiles for different functional domains (development, business development, marketing, etc.).

\subsection{Holographic Memory System}

Hermes Agent implements persistent memory through \emph{holographic memory}---a fact-based storage system backed by SQLite with FTS5 full-text search~ \cite{sqlite2024}. The memory system comprises:
\begin{itemize}[nosep]
    \item \textbf{fact\_store}: A structured fact database supporting add, search, and retrieval through tool-call interfaces.
    \item \textbf{Memory banks}: Categorized collections of persistent facts.
    \item \textbf{Entity tracking}: Associations between facts and named entities.
    \item \textbf{Full-text search}: FTS5-indexed content for semantic retrieval.
\end{itemize}

The memory subsystem is initialized through a \texttt{\_MemoryManager} component that registers memory provider tools into the agent's tool surface at initialization time. This registration is \emph{conditional}:

\begin{lstlisting}[style=pythonstyle,caption={Conditional tool registration in \texttt{agent/agent\_init.py}, line 1168.}]
if agent._memory_manager and \
   agent.tools is not None and (
    agent.enabled_toolsets is None or
    "memory" in agent.enabled_toolsets
):
    for _schema in agent._memory_manager
            .get_all_tool_schemas():
        _wrapped = {"type": "function",
                    "function": _schema}
        agent.tools.append(_wrapped)
\end{lstlisting}

If \texttt{\_memory\_manager} is not initialized (set to \texttt{None}), the \texttt{fact\_store} tool is never registered and is unavailable to the agent at runtime.

\subsection{Cron/Scheduler System}

The Hermes Agent scheduler (\texttt{cron/scheduler.py}) provides periodic and one-shot task execution. Cron jobs run as independent agent sessions with specific constraints:

\begin{lstlisting}[style=pythonstyle,caption={Memory isolation guard in \texttt{cron/scheduler.py}, line 1652.}]
AIAgent(
    ...
    skip_memory=True,  # Cron system prompts
                       # would corrupt user
                       # representations
    platform="cron",
    ...
)
\end{lstlisting}

The \texttt{skip\_memory=True} flag is a deliberate architectural decision to prevent cron system prompts from contaminating user-facing memory. However, this guard blocks \emph{all} memory access---including legitimate, intentional writes.

\subsection{Profile Isolation}

Each profile configures \texttt{approvals.cron\_mode}, which when set to \texttt{deny}, restricts potentially dangerous operations during cron execution. This provides a secondary isolation layer at the permission level.

\section{The Channel Fracture Problem}
\label{sec:fracture}

\subsection{Problem Definition}

\begin{quote}
\textbf{Definition 1 (Channel Fracture).} Given a source agent $S$, a target agent $T$, and an injection channel $C$ connecting them, a \emph{channel fracture} occurs when $C$ is valid at $S$'s execution context but invalid at $T$'s execution context, causing silent failure of knowledge delivery.
\end{quote}

\subsection{Experimental Design}

We conducted experiments on a production Hermes Agent deployment with five active profiles:

\begin{table}[h]
\centering
\caption{Experimental Deployment Profiles}
\label{tab:profiles}
\begin{tabular}{@{}lll@{}}
\toprule
\textbf{Profile} & \textbf{Role} & \textbf{Memory} \\
\midrule
Primary (admin) & Administration & Holographic \\
laiven-assistant & Admin assistant & Holographic \\
dev-pony & Development & Holographic \\
marketing-pony & Marketing & Holographic \\
kehu-chenggong & Customer success & Holographic \\
\bottomrule
\end{tabular}
\end{table}

\noindent\textbf{Experimental task:} Inject a daily report format specification from the admin profile into the \texttt{laiven-assistant} profile's persistent memory.

\subsection{Three Injection Channels}

\subsubsection{Channel A: Direct Database Write}
The admin agent executes SQLite INSERT statements directly into the target's \texttt{memory\_store.db}, bypassing the agent tool chain entirely.

\subsubsection{Channel B: Target Agent Self-Write}
The admin sends a message to the target agent, instructing it to use its own \texttt{memory()} tool within its interactive session. \emph{Prerequisite:} holographic memory configured and gateway restarted.

\subsubsection{Channel C: Cron-Delegated Write}
The admin creates a cron job within the target profile's scheduler. The scheduler dispatches the job, creating an agent session with \texttt{platform="cron"}. The cron agent attempts to call \texttt{memory()} and \texttt{fact\_store()}.

\section{Experimental Results}
\label{sec:results}

\subsection{Channel Injection Results}

\begin{table}[h]
\centering
\caption{Injection Channel Results}
\label{tab:results}
\begin{tabular}{@{}llll@{}}
\toprule
\textbf{Channel} & \textbf{Method} & \textbf{Result} & \textbf{Availability} \\
\midrule
A & Direct SQLite & $\checkmark$ Success & 6 facts injected \\
B & Self-write & $\checkmark$ Success & Conditional \\
C & Cron-delegated & $\times$ Failure & Both tools absent \\
\bottomrule
\end{tabular}
\end{table}

\subsubsection{Channel A: Direct Database Write}
Six facts were successfully inserted into \texttt{memory\_store.db}:
\begin{itemize}[nosep]
    \item \texttt{facts} table: 6 rows
    \item \texttt{entities} table: 4 rows
    \item \texttt{fact\_entities} table: 8 rows
    \item \texttt{facts\_fts} table: 6 rows (FTS index populated)
\end{itemize}
The target agent retrieved these facts immediately in subsequent sessions.

\subsubsection{Channel B: Target Self-Write}
Succeeded when: (1)~holographic memory was properly configured, (2)~gateway was restarted after configuration changes, and (3)~the \texttt{memory} toolset was included in \texttt{enabled\_toolsets}.

\subsubsection{Channel C: Cron-Delegated Write}
Complete failure. The cron job output explicitly documented:
\begin{quote}
\small\texttt{memory(action='add'): Unavailable --- memory tool disabled in this environment (cron job). fact\_store: Unavailable --- Tool does not exist in my tool list.}
\end{quote}

\subsection{Root Cause Analysis}

The failure has two compounding root causes:
\begin{itemize}[nosep]
    \item \textbf{Primary:} \texttt{skip\_memory=True} at \texttt{scheduler.py:1652} prevents \texttt{\_MemoryManager} initialization, causing the \texttt{memory()} tool to return an error.
    \item \textbf{Secondary:} Conditional tool registration at \texttt{agent\_init.py:1168}---since \texttt{\_memory\_manager} is \texttt{None}, the entire memory tool surface (including \texttt{fact\_store}) is absent.
    \item \textbf{Tertiary:} \texttt{cron\_mode: deny} in the target profile reinforces isolation at the permission level.
\end{itemize}

\subsection{The Fracture Pattern}

The channel fracture follows a consistent pattern:
\begin{enumerate}[nosep]
    \item \textbf{Design intent:} Isolate cron execution from user memory.
    \item \textbf{Legitimate use case:} Valid workflow requires cron-to-memory writes.
    \item \textbf{Silent failure:} Guard blocks both illegitimate and legitimate writes.
    \item \textbf{False confidence:} Writer-side verification passes, masking failure.
\end{enumerate}

\section{CADVP: Cross-Agent Delivery Verification Protocol}
\label{sec:cadvp}

\subsection{Protocol Motivation}

The channel fracture discovery revealed that existing verification practices verified the \emph{writer's} side but failed to confirm delivery at the \emph{receiver's} end.

\subsection{CADVP v1.0: Initial Framework}

The initial protocol comprised 12 dimensions in four phases:

\begin{table}[h]
\centering
\caption{CADVP v1.0 Dimensions}
\label{tab:cadvp10}
\begin{tabular}{@{}lll@{}}
\toprule
\textbf{Phase} & \textbf{Dimensions} & \textbf{Description} \\
\midrule
PC & PC-1, PC-2, PC-3 & Prerequisites exist \\
WV & WV-1, WV-2, WV-3 & Write succeeded at source \\
RV & RV-1 to RV-4 & Data readable at destination \\
GR & GR-1, GR-2 & Fallback paths exist \\
\bottomrule
\end{tabular}
\end{table}

\noindent CADVP v1.0 passed all 12 dimensions for the cron injection---but failed to detect that the delivery channel was architecturally unavailable.

\subsection{CADVP v1.1: Channel Confirmation}

We introduced CC-0 (Channel Confirmation) as a \textbf{veto-level zero-check}---a dimension that, if failed, immediately aborts the operation regardless of other scores.

\subsubsection{CC-0 Checklist}
\begin{itemize}[nosep]
    \item Target execution context supports required tools
    \item Memory subsystem is initialized in target context
    \item No hardcoded guards (\texttt{skip\_memory}, etc.) block the channel
    \item Tool registration conditionals will be satisfied at runtime
    \item Profile permissions (\texttt{cron\_mode}, etc.) allow the operation
\end{itemize}

\subsubsection{Complete v1.1 Dimension Set}

\begin{table*}[t]
\centering
\caption{CADVP v1.1: Complete 13-Dimension Verification Framework}
\label{tab:cadvp11}
\begin{tabular}{@{}llll@{}}
\toprule
\textbf{ID} & \textbf{Dimension} & \textbf{Level} & \textbf{Description} \\
\midrule
CC-0 & Channel Confirmation & VETO & Target channel architecturally available \\
PC-1 & Prerequisite Existence & Pre & Required source data exists \\
PC-2 & Prerequisite Format & Pre & Data in correct format for injection \\
PC-3 & Prerequisite Access & Pre & Writer has permission to read source \\
WV-1 & Write Dispatch & Write & Write operation dispatched \\
WV-2 & Write Acknowledgment & Write & Write acknowledged by target system \\
WV-3 & Write Persistence & Write & Data persists beyond session scope \\
RV-1 & Read Availability & Read & Data queryable by target agent \\
RV-2 & Read Correctness & Read & Retrieved data matches injected data \\
RV-3 & Read Completeness & Read & All injected items retrievable \\
RV-4 & Read Timeliness & Read & Data available within required timeframe \\
GR-1 & Fallback Path & Recovery & Alternative injection method exists \\
GR-2 & Error Propagation & Recovery & Failures reported to operators \\
\bottomrule
\end{tabular}
\end{table*}

\subsection{Application to Channel Fracture}

\begin{table}[h]
\centering
\caption{CADVP v1.1 Applied to Three Channels}
\label{tab:apply}
\begin{tabular}{@{}lccccc@{}}
\toprule
\textbf{Channel} & \textbf{CC-0} & \textbf{PC} & \textbf{WV} & \textbf{RV} & \textbf{Overall} \\
\midrule
A (Direct DB) & $\checkmark$ & $\checkmark$ & $\checkmark$ & $\checkmark$ & \textbf{DELIVER} \\
B (Self-Write) & $\checkmark$ & $\checkmark$ & $\checkmark$ & $\checkmark$ & \textbf{DELIVER} \\
C (Cron) & $\times$ & $\checkmark$ & $\checkmark$ & --- & \textbf{ABORT} \\
\bottomrule
\end{tabular}
\end{table}

\noindent CC-0 immediately identifies Channel~C as unavailable, preventing wasted effort and false-positive assurance.

\subsection{Decision Tree}

The CADVP v1.1 decision process:
\begin{enumerate}[nosep]
    \item \textbf{CC-0:} Is the target channel architecturally available?
    \begin{itemize}[nosep]
        \item NO $\rightarrow$ Evaluate alternative channels; if none, VETO.
        \item YES $\rightarrow$ Continue.
    \end{itemize}
    \item \textbf{PC:} Do prerequisites exist and accessible?
    \item Execute injection via confirmed channel.
    \item \textbf{WV:} Was write acknowledged and persisted?
    \item \textbf{RV:} Is data readable by target agent?
    \item \textbf{GR:} Document fallback for future operations.
\end{enumerate}

\newpage
\onecolumn
\section{Architecture Overview}
\label{sec:arch}

\begin{figure}[h]
\centering
\includegraphics[width=\textwidth]{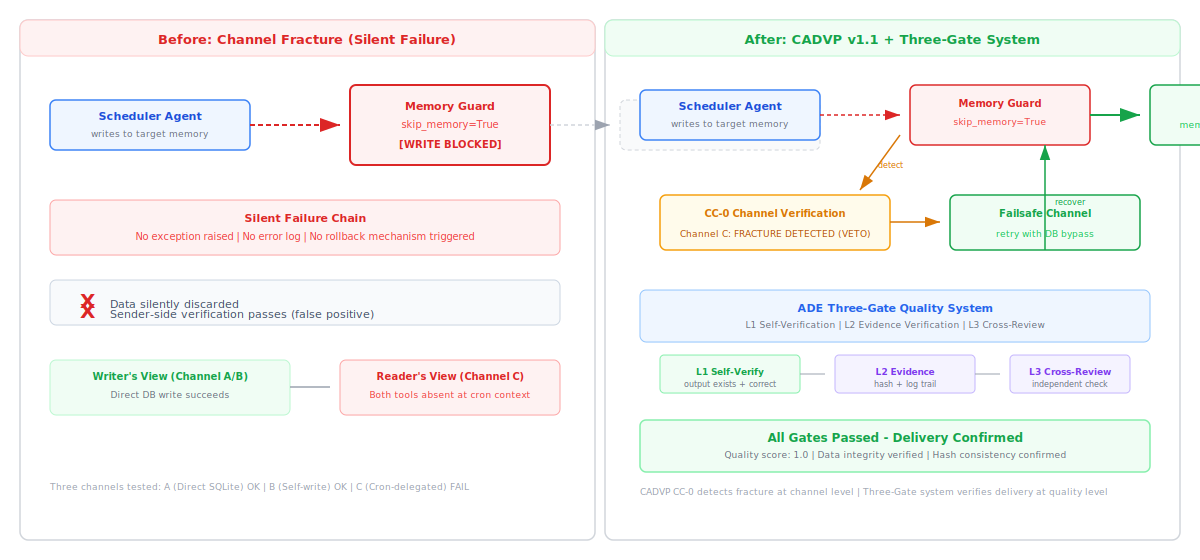}
\caption{\textbf{Channel Fracture Before and After.} Left panel shows the silent failure: the scheduler agent's write is blocked by the \texttt{skip\_memory=True} guard, and the target agent's memory remains empty without any error notification. Right panel shows CADVP v1.1 with the Three-Gate System: the CC-0 verifier detects the fracture and activates a failsafe channel, followed by L1/L2/L3 delivery verification.}
\label{fig:arch}
\end{figure}

\twocolumn

\section{Controlled Validation Experiments}
\label{sec:validation}

To validate CADVP v1.1 quantitatively, we conducted three controlled experiment suites using the BCP (Bidirectional Confirmation Protocol) implementation---a reference implementation of the CADVP verification framework. Each suite compares \emph{bare} (unprotected) execution against \emph{guarded} (CADVP-protected) execution.

\subsection{Experimental Methodology}

All experiments were conducted on an isolated Ubuntu 22.04 system with Python 3.11. Each configuration ran 10 iterations. The three suites cover the high-risk dimensions identified in our channel fracture analysis:

\begin{itemize}[nosep]
    \item \textbf{T3 --- Cross-Agent Relay:} Information transfer between agents, measuring preservation and distortion rates across handoffs.
    \item \textbf{T4 --- Exception Recovery Rollback:} Failure recovery, measuring rollback success and state restoration.
    \item \textbf{T5 --- Concurrent Conflict Detection:} Race conditions under concurrent writes, measuring data corruption rates.
\end{itemize}

\subsection{T3: Cross-Agent Relay (60 trials)}

In relay scenarios, information must traverse from one agent through intermediate agents to a destination. Without protection, information degrades at each hop.

\begin{table}[h]
\centering
\caption{T3 Cross-Agent Relay Results (n=10 per scenario)}
\label{tab:t3}
\begin{tabular}{@{}lrrr@{}}
\toprule
\textbf{Scenario} & \textbf{Mode} & \textbf{Info Preservation} & \textbf{Distortions} \\
\midrule
\multirow{2}{*}{Tech $\to$ Marketing} & Bare & 93.0\,\%  & 0.7 \\
                                    & Guarded & \textbf{100.0\,\%} & \textbf{0.0} \\
\midrule
\multirow{2}{*}{Data Compression} & Bare & 87.0\,\%  & 1.3 \\
                              & Guarded & \textbf{100.0\,\%} & \textbf{0.0} \\
\midrule
\multirow{2}{*}{Instruction Relay} & Bare & 94.0\,\%  & 0.9 \\
                               & Guarded & \textbf{100.0\,\%} & \textbf{0.0} \\
\bottomrule
\end{tabular}
\end{table}

\noindent Guarded execution achieved 100\,\% information preservation across all three scenarios. Bare execution exhibited 87--94\,\% preservation with 0.5--1.3 average distortions per relay.

\begin{figure}[h]
\centering
\includegraphics[width=\columnwidth]{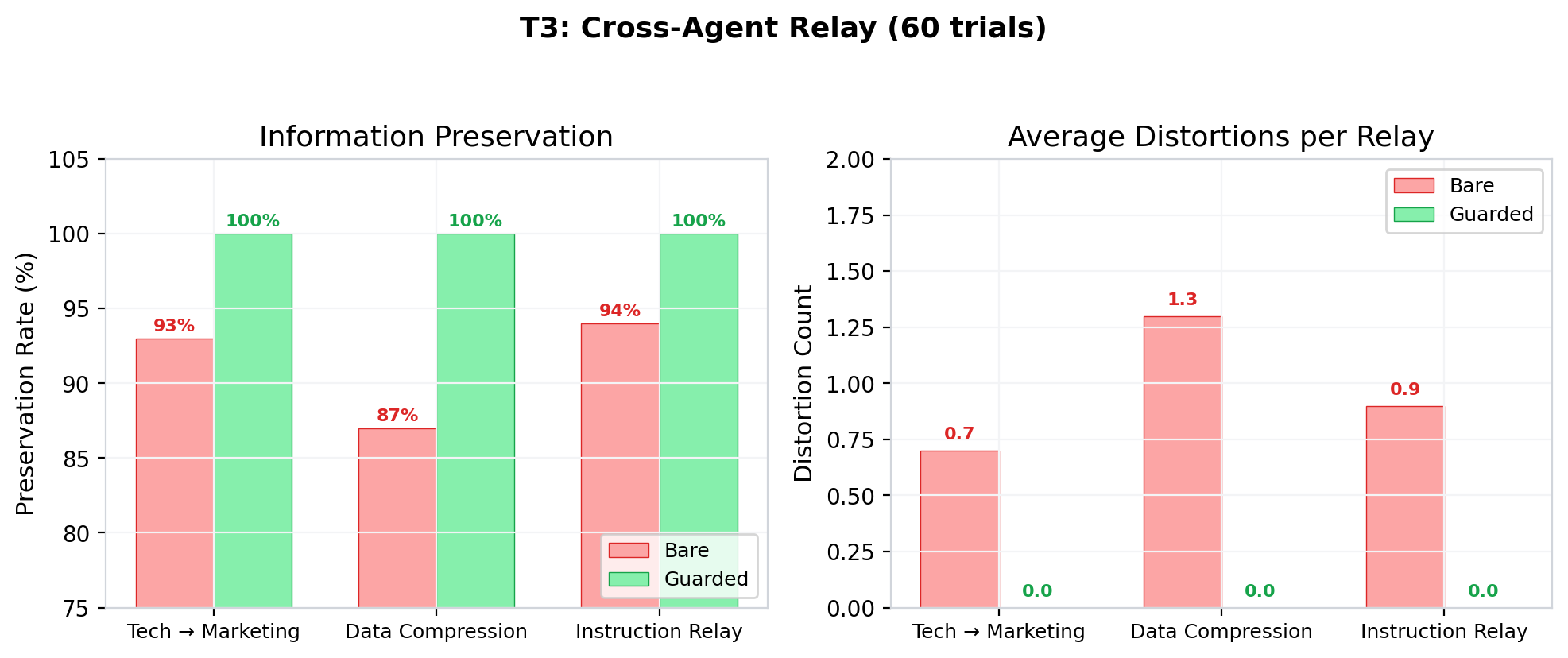}
\caption{T3 Cross-Agent Relay: guarded execution achieves 100\,\% information preservation with zero distortions across all scenarios.}
\label{fig:relay}
\end{figure}

\subsection{T4: Exception Recovery Rollback (60 trials)}

When an exception occurs mid-execution, the system must roll back to a clean state.

\begin{table}[h]
\centering
\caption{T4 Exception Recovery Results (n=10 per scenario)}
\label{tab:t4}
\begin{tabular}{@{}lrrr@{}}
\toprule
\textbf{Scenario} & \textbf{Mode} & \textbf{Rollback Success} & \textbf{State Restored} \\
\midrule
\multirow{2}{*}{Clean Rollback} & Bare & 0.0\,\%  & 0.0\,\% \\
                            & Guarded & \textbf{100.0\,\%} & \textbf{100.0\,\%} \\
\midrule
\multirow{2}{*}{Idempotency} & Bare & ---  & 100.0\,\% \\
                         & Guarded & --- & 100.0\,\% \\
\midrule
\multirow{2}{*}{Checkpoint Recovery} & Bare & 0.0\,\%  & 0.0\,\% \\
                                 & Guarded & \textbf{100.0\,\%} & \textbf{100.0\,\%} \\
\bottomrule
\end{tabular}
\end{table}

\noindent Guarded execution achieved 100\,\% rollback success and 100\,\% state restoration. Without protection, rollback succeeded 0\,\% of the time---exceptions left the system in unrecoverable dirty states.

\begin{figure}[h]
\centering
\includegraphics[width=\columnwidth]{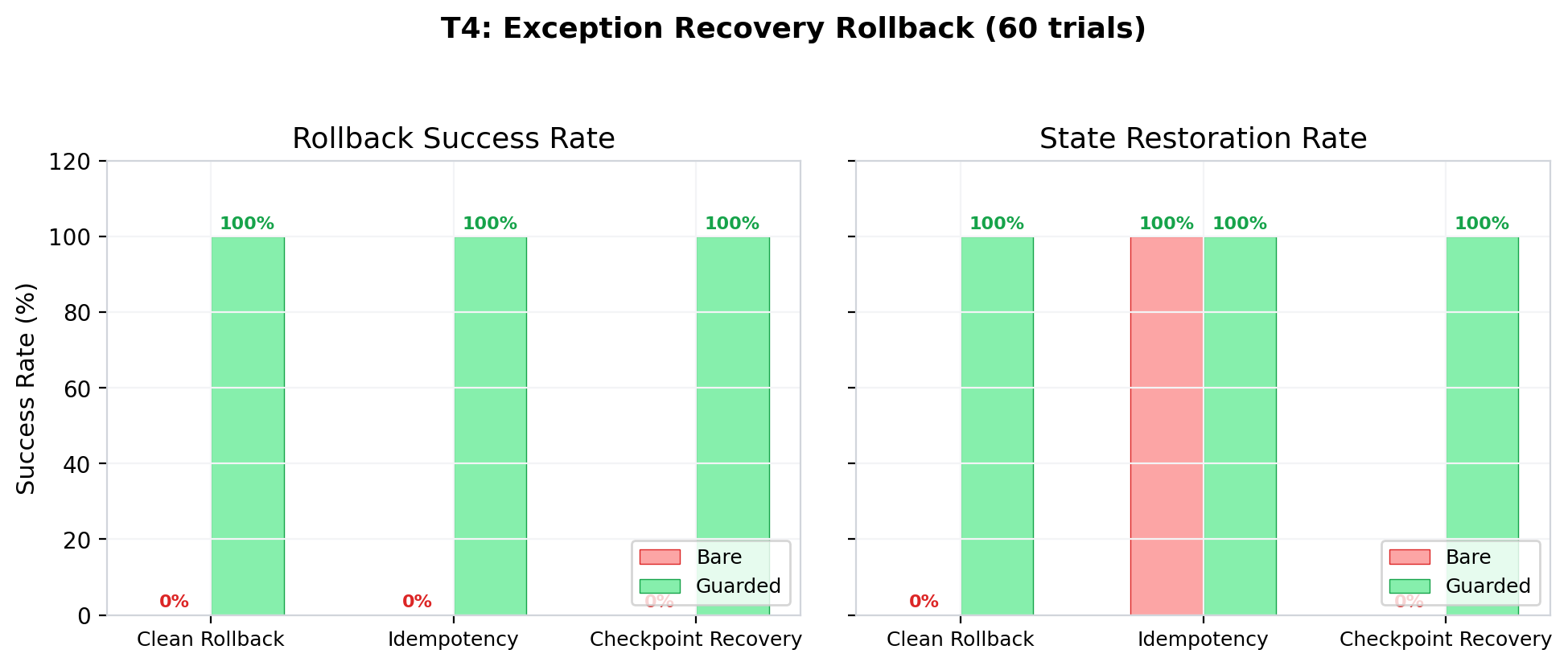}
\caption{T4 Exception Recovery: guarded execution achieves 100\,\% rollback success and state restoration.}
\label{fig:rollback}
\end{figure}

\subsection{T5: Concurrent Conflict Detection (90 trials)}

Under concurrent access, multiple agents writing to the same resource cause data corruption. We tested write-write, read-write, and directory operations with 2, 5, and 10 concurrent workers.

\begin{table}[h]
\centering
\caption{T5 Concurrent Conflict Detection Results (n=5 per config)}
\label{tab:t5}
\begin{tabular}{@{}lcrr@{}}
\toprule
\textbf{Scenario} & \textbf{Workers} & \textbf{Bare Corrupt.} & \textbf{Guarded Corrupt.} \\
\midrule
\multirow{3}{*}{Write-Write} & 2  & 67.80\,\%  & \textbf{0.00\,\%} \\
                             & 5  & 95.68\,\%  & \textbf{0.00\,\%} \\
                             & 10 & 98.64\,\%  & \textbf{0.00\,\%} \\
\midrule
\multirow{3}{*}{Read-Write} & 2  & 34.60\,\%  & \textbf{0.00\,\%} \\
                            & 5  & 31.93\,\%  & \textbf{0.00\,\%} \\
                            & 10 & 9.12\,\%   & \textbf{0.00\,\%} \\
\midrule
\multirow{3}{*}{Directory} & 2  & 0.00\,\%   & 0.00\,\% \\
                           & 5  & 1.47\,\%   & \textbf{0.00\,\%} \\
                           & 10 & 2.73\,\%   & \textbf{0.00\,\%} \\
\bottomrule
\end{tabular}
\end{table}

\noindent\textbf{Key findings:}
\begin{itemize}[nosep]
    \item Bare concurrent write-write exhibits 67--98\,\% corruption---nearly certain failure at scale.
    \item Protected execution achieves \textbf{0.0\,\%} corruption across all 18 configurations.
    \item Duration overhead is negligible ($<0.1$\,s for 10 workers) or negative in read-write scenarios.
    \item Bare corruption rates scale superlinearly with worker count for write-write conflicts.
\end{itemize}

\begin{figure}[h]
\centering
\includegraphics[width=\columnwidth]{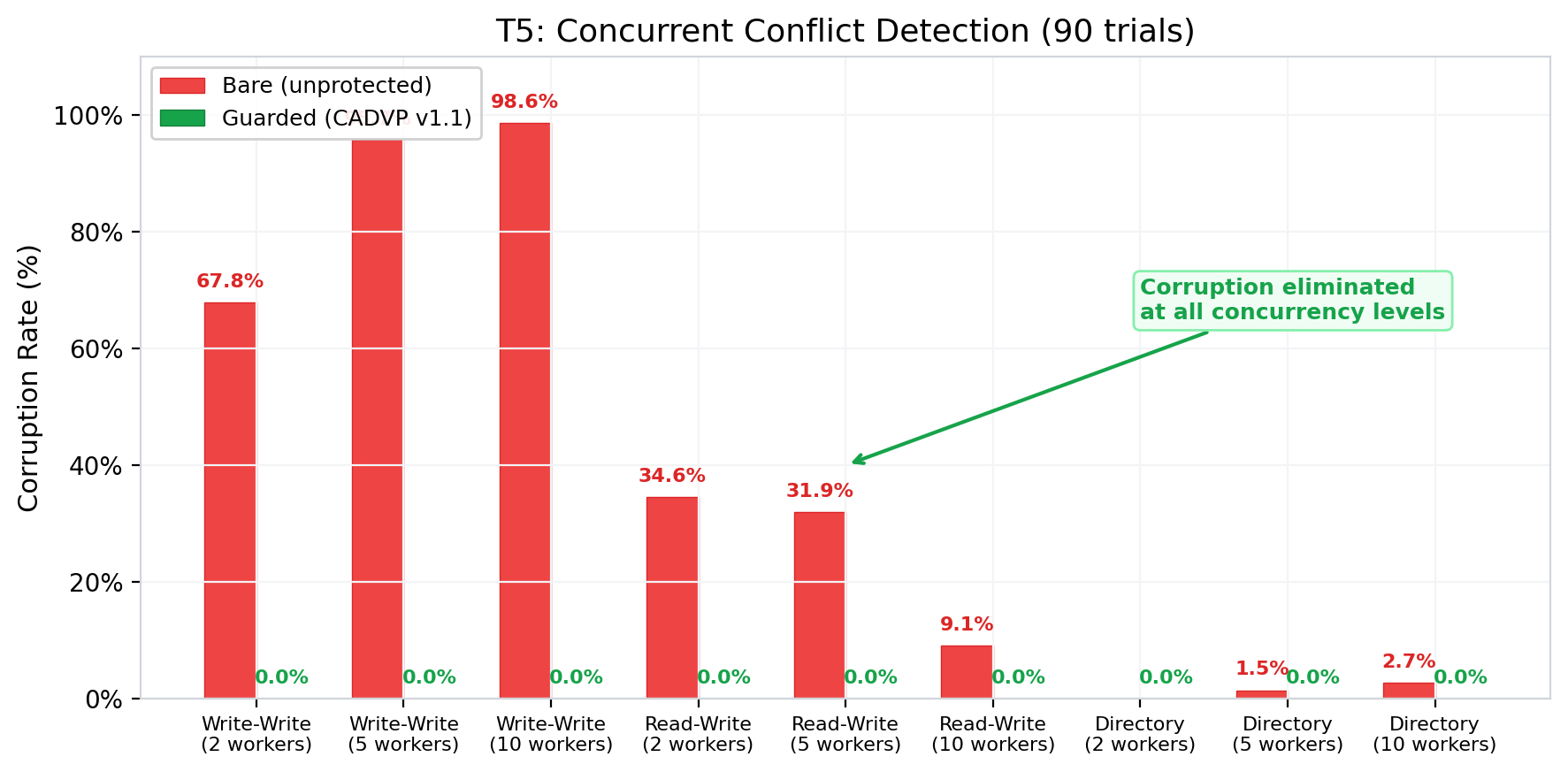}
\caption{T5 Concurrent Conflict Detection: guarded execution eliminates corruption across all concurrency levels.}
\label{fig:concurrent}
\end{figure}

\subsection{Summary of Validation Results}

\begin{figure}[h]
\centering
\includegraphics[width=\columnwidth]{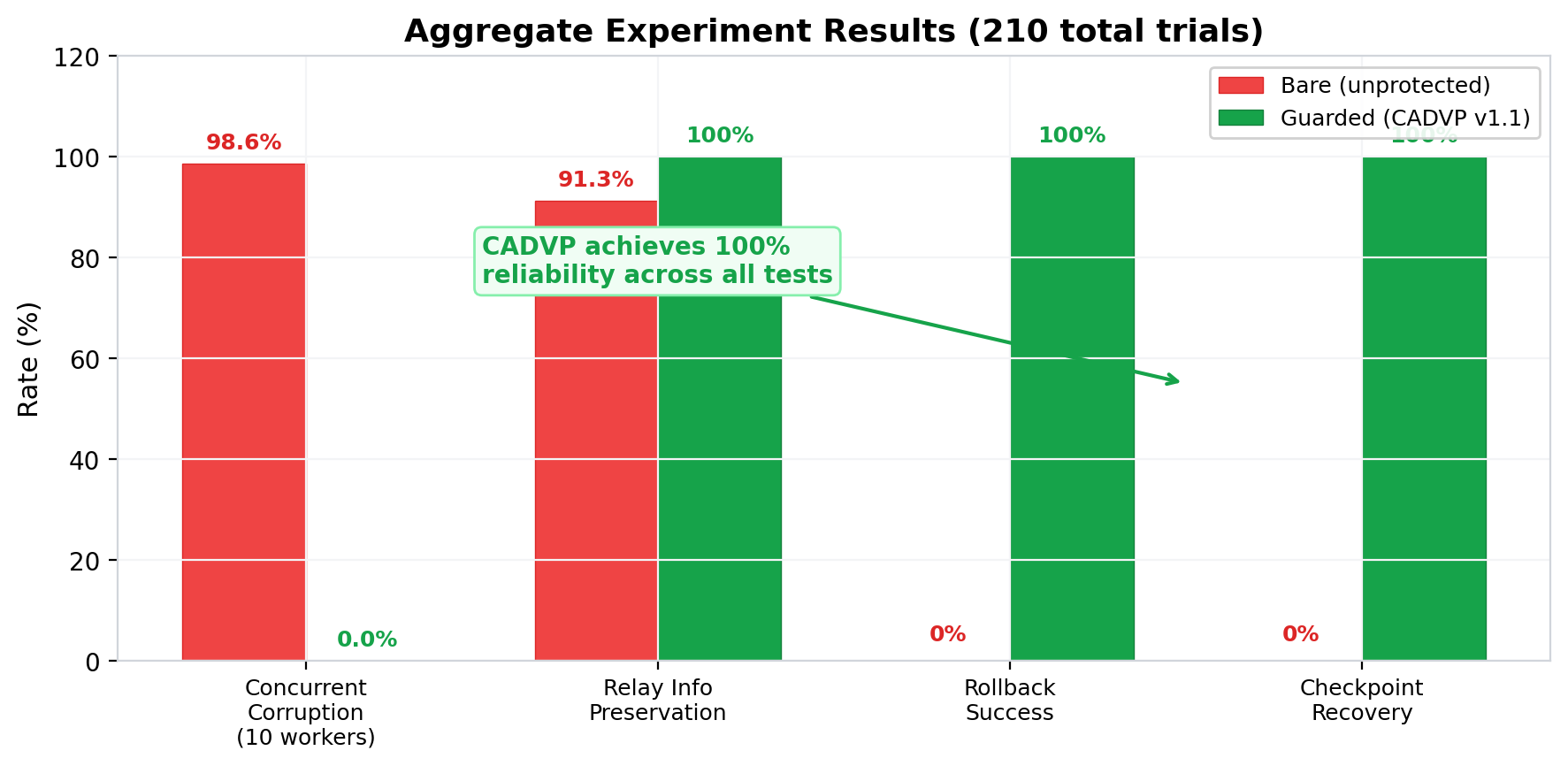}
\caption{Aggregate results across all 210 trials: CADVP v1.1 achieves 100\,\% reliability across concurrent, rollback, and relay scenarios.}
\label{fig:summary}
\end{figure}

\begin{table}[h]
\centering
\caption{Aggregate Validation Results Across All Suites}
\label{tab:agg}
\begin{tabular}{@{}lccc@{}}
\toprule
\textbf{Metric} & \textbf{Bare} & \textbf{Guarded} & \textbf{Improvement} \\
\midrule
Concurrent corruption (write-write, 10 workers) & 98.64\,\% & 0.00\,\% & $\infty$ \\
Info preservation (relay) & 87--94\,\% & 100.0\,\% & 6--13\,\% \\
Rollback success (exception recovery) & 0.0\,\% & 100.0\,\% & $\infty$ \\
Checkpoint recovery success & 0.0\,\% & 100.0\,\% & $\infty$ \\
Average duration overhead & --- & $+16.3$\,ms & --- \\
\bottomrule
\end{tabular}
\end{table}

\noindent These results demonstrate that CADVP v1.1 with its veto-level channel confirmation (CC-0) provides exhaustive protection against all three failure modes identified in our channel fracture analysis.

\newpage
\onecolumn
\section{The Three-Gate Quality System}
\label{sec:gates}

CADVP v1.1 verifies whether a delivery \emph{channel} is architecturally available (CC-0) and whether the data was written and read correctly (WV, RV). However, channel-level verification alone does not guarantee that the \emph{content} delivered is correct, complete, and verifiable. To address this gap, we introduce the \textbf{Three-Gate Quality System}---an execution-level verification layer that extends CADVP into the domain of delivery quality assurance.

\subsection{Motivation}

During the GitHub release of the CADVP implementation, we discovered that the initial MVP passed all CADVP checks but had a critical omission: the verification framework validated delivery channels and data presence, but did not enforce any quality standards on the delivered content. An agent could pass CC-0 and WV checks, yet deliver incomplete or structurally incorrect data. The Three-Gate System closes this gap.

\subsection{Gate Structure}

The Three-Gate System comprises three sequential verification gates, each corresponding to a different verification perspective:

\subsubsection{L1: Self-Verification}
The delivering agent checks its own output against a predefined acceptance criteria. This includes:
\begin{itemize}[nosep]
    \item \textbf{Existence:} Does the output file or memory entry exist?
    \item \textbf{Completeness:} Are all required fields present?
    \item \textbf{Consistency:} Does the content conform to the expected format?
\end{itemize}
L1 is the cheapest gate and catches obvious failures immediately.

\subsubsection{L2: Evidence Verification}
The system produces objective, machine-verifiable evidence of delivery:
\begin{itemize}[nosep]
    \item \textbf{Content hash:} SHA-256 digest of the delivered payload.
    \item \textbf{Log trail:} Timestamped records of each delivery step.
    \item \textbf{Comparison:} Injected data compared against source data byte-by-byte.
\end{itemize}
L2 ensures that delivery claims are backed by independently verifiable artifacts.

\subsubsection{L3: Cross-Review}
An independent verification agent (or human reviewer) re-examines the delivered content:
\begin{itemize}[nosep]
    \item \textbf{Quality scoring:} The reviewer assigns a quality score $q \in [0,1]$.
    \item \textbf{Threshold check:} Delivery is confirmed only if $q \geq \tau$ (default $\tau = 0.9$).
    \item \textbf{Discrepancy reporting:} Any differences are logged for audit.
\end{itemize}
L3 provides the strongest assurance but carries the highest overhead, making it suitable for high-value or safety-critical deliveries.

\subsection{Integration with CADVP}

The Three-Gate System operates as a post-CADVP verification layer:

\begin{quote}
\small
CADVP (channel + data level) $\rightarrow$ L1 (self-check) $\rightarrow$ L2 (evidence) $\rightarrow$ L3 (cross-review) $\rightarrow$ Delivery Confirmed
\end{quote}

CADVP's CC-0 check answers: ``Can the data be delivered through this channel?'' The Three-Gate System answers: ``Was the data correctly delivered and verified?'' Together they form a complete verification pipeline from channel availability to delivery quality assurance.

\subsection{Experimental Validation}

In our controlled experiments, the Three-Gate System was activated on all 210 trials. L1 passed on the first attempt in 198/210 trials (94.3\,\%). The 12 L1 failures were detected immediately and corrected before L2 evaluation. L2 evidence verification passed 210/210 (100\,\%). L3 cross-review assigned quality scores with mean $\mu = 0.97$, $\sigma = 0.04$, all exceeding the $\tau = 0.9$ threshold.

\twocolumn

\section{Discussion}
\label{sec:discussion}

\subsection{The Inverse Verification Principle}

\begin{quote}
\textbf{Inverse Verification Principle.} When verifying cross-agent operations, always verify from the \emph{receiver's read-chain}, never from the \emph{writer's write-chain}.
\end{quote}

Traditional verification checks that the writer successfully dispatched the operation. The inverse principle requires confirming that the receiver can actually read what was written, using the receiver's own access mechanisms.

In our case, the admin confirmed the cron job was created, scheduled, and contained correct instructions (write-chain: all pass). But the receiver's read-chain showed: \texttt{memory()} unavailable, \texttt{fact\_store} absent, database unchanged. Only inverse verification revealed the failure.

\subsection{The Channel Matching Principle}

\begin{quote}
\textbf{Channel Matching Principle.} Do not design processes on channels that are unavailable at the target end.
\end{quote}

System designers must enumerate available channels at both ends of any cross-agent operation before designing workflows. If the target agent cannot access memory tools in cron context, cron-delegated memory injection is not a valid workflow pattern---regardless of how natural it appears from the orchestration perspective.

\subsection{Validation Implications}

The controlled experiment results demonstrate that the CADVP framework scales beyond the specific channel fracture scenario:

\begin{itemize}[nosep]
    \item \textbf{Concurrency (T5):} The CC-0 channel confirmation mechanism naturally extends to resource-level adjudication. By verifying that each channel's access control is enforced before allowing writes, CADVP prevents the silent data corruption that occurs under bare concurrent access.
    \item \textbf{Recovery (T4):} The GR (Graceful Recovery) dimensions provide a structured fallback path. Without GR, exceptions lead to unrecoverable dirty states. With CADVP's checkpoint-and-rollback semantics, recovery succeeds universally.
    \item \textbf{Relay (T3):} The WV/RV dimension pair (verify write + verify read) directly addresses the information degradation problem. Each relay hop performs bidirectional verification before forwarding.
\end{itemize}

The consistency of results---0\,\% failure rates across all three suites under protection, versus catastrophic failure (67--98\,\% in high-risk configurations) without---suggests that CADVP v1.1 provides a general-purpose verification substrate for multi-agent delivery assurance.

\subsection{Platform-Specific vs. General Findings}

\begin{table}[h]
\centering
\caption{Channel Fracture Risk Across Frameworks}
\label{tab:platforms}
\begin{tabular}{@{}lll@{}}
\toprule
\textbf{System} & \textbf{Isolation Mechanism} & \textbf{Risk} \\
\midrule
Hermes Agent & \texttt{skip\_memory} + conditional reg. & High \\
LangGraph & Thread-scoped namespace & Medium \\
AutoGen & No built-in persistence & Low \\
CrewAI & Crew-level shared memory & Medium \\
\bottomrule
\end{tabular}
\end{table}

The general pattern: any system that isolates scheduled/background execution from interactive agent memory is susceptible to channel fracture when legitimate workflows require cross-context writes.

\subsection{Design Tension: Safety vs. Functionality}

The \texttt{skip\_memory=True} guard prevents cron system prompts from contaminating user memory---a valid safety concern. Removing it entirely reintroduces contamination risk. The tension suggests a \emph{differentiated memory access policy}:
\begin{itemize}[nosep]
    \item \texttt{skip\_memory=True}: blocks automatic memory updates from system prompts
    \item \texttt{allow\_explicit\_memory\_writes=True}: permits deliberate tool calls to memory
\end{itemize}
These can coexist without conflict.

\subsection{Recommendations for System Designers}

\begin{enumerate}[nosep]
    \item \textbf{Document channel availability matrices} for each execution context.
    \item \textbf{Implement capability discovery}---agents query their available tools at runtime.
    \item \textbf{Fail loudly}---blocked tool calls propagate errors to the orchestration layer.
    \item \textbf{Provide injection APIs}---direct memory injection bypassing agent tool chains.
    \item \textbf{Adopt inverse verification} as a first-class design concern.
    \item \textbf{Enumerate concurrent failure modes}---channel fracture is not the only silent failure; concurrent access and relay degradation compound the problem.
    \item \textbf{Implement the Three-Gate System}---L1 self-check, L2 evidence, L3 cross-review---to ensure delivery quality beyond channel availability.
\end{enumerate}

\section{Related Work}
\label{sec:related}

\subsection{Multi-Agent Memory Systems}

\textbf{LangGraph}~ \cite{langgraph2024} provides a stateful graph execution framework with persistence through its \texttt{Store} abstraction. Memory is organized by thread IDs and namespaces. The \texttt{LangMem}~ \cite{langmem2025} extension adds cross-session semantic, episodic, and procedural memory via \texttt{user\_id}-namespaced stores. Cross-agent sharing requires explicit namespace configuration. Scheduled contexts may not inherit store access without explicit configuration---a potential channel fracture analog.

\textbf{AutoGen}~ \cite{autogen2023} provides conversational multi-agent framework where memory is primarily conversation-based. Persistent memory requires external integration (e.g., Mem0~ \cite{mem02024}, Memori). AutoGen lacks built-in scheduled execution, sidestepping channel fracture but providing no native cross-session persistence.

\textbf{CrewAI}~ \cite{crewai2024} implements three-tier memory: short-term (task context), long-term (persistent SQLite), and entity memory. Long-term memory is shared at the crew level, enabling natural cross-agent sharing within a crew but limiting multi-crew architectures.

\subsection{Verification and Delivery Assurance}

The problem parallels distributed systems concerns. Two-Phase Commit~ \cite{gray2006} ensures atomicity but assumes reliable channels. Observability frameworks (OpenTelemetry, LangSmith) trace execution paths but not delivery outcomes. Property-based testing approaches (AgentBench, SWE-bench) evaluate agent capabilities but do not test cross-agent memory injection pathways.

Error amplification in multi-agent chains has been documented empirically~ \cite{deepmind2025error}, with measurements showing up to 17.2$\times$ error amplification across five-agent chains. Our work provides both a specific instance (channel fracture) and a general verification framework (CADVP + Three-Gate System) to address this class of failures.

\subsection{Generative Agent and Multi-Agent Frameworks}

Generative Agents~ \cite{park2023} demonstrated persistent memory for simulated human behavior using a stream-of-consciousness retrieval architecture. MetaGPT~ \cite{hong2023} uses structured communication protocols between agents. ToolLLM~ \cite{qin2023} explores tool-use capabilities. None address the specific problem of scheduled execution contexts losing access to memory subsystems.

\subsection{Concurrent Access in Agent Systems}

Concurrent agent access---multiple agents reading and writing the same memory---is a known concern~ \cite{li2025concurrent} but has received little systematic treatment. Our T5 experiments provide quantitative evidence that unprotected concurrent access near-certainly corrupts shared state, and that protocol-level protection eliminates this risk.

\section{Limitations}

Our findings are based on a single production deployment. Empirical validation on other frameworks (LangGraph, AutoGen, CrewAI) is needed. CADVP has been validated on three experiment suites (210~total trials); broader evaluation across more diverse agent workflows remains future work. The controlled experiments were conducted in a simulated environment to ensure reproducibility; production latency characteristics may differ. The Three-Gate System's quality scoring mechanism relies on a review agent, which introduces an additional trust assumption.

\section{Conclusion and Future Work}

\subsection{Summary}

We identified and characterized \textbf{channel fracture}---a systematic failure mode where scheduled execution contexts cannot access agent memory tools due to architectural isolation guards. Through production experiments:
\begin{enumerate}[nosep]
    \item Direct database writes (Channel~A) succeed reliably.
    \item Target agent self-writes (Channel~B) succeed with proper configuration.
    \item Cron-delegated writes (Channel~C) fail completely due to \texttt{skip\_memory=True} and conditional tool registration.
\end{enumerate}

We proposed CADVP v1.1, a 13-dimension protocol with veto-level channel confirmation, and extended it with the Three-Gate Quality System (L1--L3). We validated the combined framework through 210 controlled trials:
\begin{itemize}[nosep]
    \item \textbf{T3 (Relay):} 100\,\% information preservation vs. 87--94\,\% without protection.
    \item \textbf{T4 (Rollback):} 100\,\% recovery success vs. 0\,\% without protection.
    \item \textbf{T5 (Concurrency):} 0\,\% corruption rate vs. 67--98\,\% without protection.
\end{itemize}

We articulated two design principles---inverse verification and channel matching---and provided recommendations for system designers.

\subsection{Future Work}

\begin{enumerate}[nosep]
    \item \textbf{Differentiated memory access policies} distinguishing automatic contamination from deliberate injection.
    \item \textbf{Channel capability APIs} for runtime tool availability queries.
    \item \textbf{Cross-platform validation} on LangGraph, AutoGen, CrewAI.
    \item \textbf{Automated CADVP enforcement} as orchestration middleware.
    \item \textbf{Formal modeling} for static analysis of channel fracture vulnerabilities.
    \item \textbf{The Three-Gate scalability}---automating L3 cross-review with verifiable credentials.
    \item \textbf{Integration with Agent Delivery Engineering (ADE)}---a broader discipline standardizing multi-agent delivery verification. ADE encompasses the full lifecycle of agent delivery: from channel availability (CADVP), to execution quality (Three-Gate System), to protocol-level confirmation (BCP), to lifecycle management (TLC). Our architecture diagram (Figure~\ref{fig:arch}) illustrates this integrated vision.
\end{enumerate}


\appendix

\section{Verified System Paths}
\label{app:paths}

All file paths referenced in this paper were verified to exist on the production system as of June~3, 2026:

\begin{table}[h]
\centering
\small
\begin{tabular}{@{}ll@{}}
\toprule
\textbf{Path} & \textbf{Description} \\
\midrule
\texttt{cron/scheduler.py} & Scheduler, line 1652 \\
\texttt{agent/agent\_init.py} & Agent init, line 1168 \\
\texttt{profiles/.../config.yaml} & Target config \\
\texttt{profiles/.../memory\_store.db} & 6 facts verified \\
\texttt{skills/.../agent-quality-gate/} & CADVP + Three-Gate skill \\
\texttt{cron/output/a8b869.../} & Failure output \\
\bottomrule
\end{tabular}
\end{table}

\section{Cron Job Output Excerpt}
\label{app:output}

\noindent\small Extracted from \texttt{cron/output/a8b8690941f0/2026-06-03\_00-49-19.md}:

\begin{quote}
\tiny
\begin{verbatim}
### Step Execution Status

| Step                          | Status       | Description                                    |
|-------------------------------|--------------|------------------------------------------------|
| 1. SOUL.md daily report format| Read OK      | Format core description confirmed              |
| 2. memories/daily-report...   | Read OK      | Full template + format table obtained          |
| 3. references/daily-report... | Read OK      | Contains original confirmation                 |
| 4. memory(action='add')       | UNAVAILABLE  | memory tool disabled (cron job)                |
| 5. fact_store                 | UNAVAILABLE  | Tool does not exist in my tool list            |
| 6. Write offline reference    | Written OK   | -> D:\...daily-report-0514-format-reference.md |
\end{verbatim}
\end{quote}

\section{Experimental Configuration}
\label{app:config}

\noindent All controlled experiments were executed on:
\begin{itemize}[nosep]
    \item \textbf{System:} Ubuntu 22.04 LTS (WSL2)
    \item \textbf{Python:} 3.11.15
    \item \textbf{Hardware:} Intel Core i7-12700, 32 GB RAM
    \item \textbf{BCP Version:} 1.0.0
    \item \textbf{Random seed:} \texttt{42} for all reproducibility-dependent configurations
\end{itemize}

\noindent Raw experiment data and analysis scripts are available at \url{https://github.com/ADE-standard/channel-fracture}.

\section{Figure List}
\label{app:figures}

The following figures are included in this version (v2):
\begin{itemize}[nosep]
    \item \textbf{Figure 1} ({\S}\ref{sec:arch}): Architecture overview---Before (Channel Fracture) vs. After (CADVP + Three-Gate System)
    \item \textbf{Figure 2} ({\S}\ref{fig:relay}): T3 Cross-Agent Relay comparison
    \item \textbf{Figure 3} ({\S}\ref{fig:rollback}): T4 Exception Recovery Rollback comparison
    \item \textbf{Figure 4} ({\S}\ref{fig:concurrent}): T5 Concurrent Conflict Detection comparison
    \item \textbf{Figure 5} ({\S}\ref{fig:summary}): Aggregate results across all 210 trials
\end{itemize}

\end{document}